\documentclass[11pt]{article}

\newif\ifopre
\oprefalse

\usepackage[english]{babel}
\usepackage[utf8]{inputenc}
\usepackage{amsmath}
\usepackage{amssymb}
\usepackage{graphicx}
\usepackage{hyperref}
\usepackage{algorithm}
\usepackage{algpseudocode}
\usepackage{todonotes}
\usepackage[round]{natbib}
\usepackage{fix-cm}

\bibpunct[, ]{(}{)}{,}{a}{}{,}%

\ifopre
\DoubleSpacedXI 
\TheoremsNumberedThrough 
\ECRepeatTheorems
\EquationsNumberedThrough 
\else
\usepackage{fullpage}
\usepackage{amsthm}
\newtheorem{theorem}{Theorem}
\theoremstyle{definition}
\newtheorem{definition}{Definition}
\newtheorem{lemma}{Lemma}
\newtheorem{example}{Example}
\newtheorem{remark}{Remark}
\fi
\newtheorem{algo}[algorithm]{Algorithm}

\begin{document}

\ifopre
\else
\title{A decomposition algorithm for computing income taxes with pass-through entities and its application to the Chilean case}
\author{Javiera Barrera\\
Facultad de Ingeniería y Ciencias\\
Universidad Adolfo Ibáñez
\and
Eduardo Moreno\\
Facultad de Ingeniería y Ciencias\\
Universidad Adolfo Ibáñez
\and
Sebastián Varas K.\\
CIRIC - INRIA Chile
\date{April 7, 2016}
}
\maketitle

\begin{abstract}
Income tax systems with ``pass-through'' entities transfer a firm’s incomes to the shareholders, which are taxed individually. In 2014, a Chilean tax reform introduced this type of entity and changed to an accrual basis that distributes incomes (but not losses) to shareholders. A crucial step for the Chilean taxation authority is to compute the final income of each individual, given the complex network of corporations and companies, usually including cycles between them.
In this paper, we show the mathematical conceptualization and the solution to the problem, proving that there is only one way to distribute incomes to taxpayers. Using the theory of absorbing Markov chains, we define a mathematical model for computing the taxable incomes of each taxpayer, and we propose a decomposition algorithm for this problem. This allows us to compute the solution accurately and with the efficient use of computational resources. Finally, we present some characteristics of the Chilean taxpayers' network and computational results of the algorithm using this network.
\end{abstract}
\fi

\section{Introduction}
\label{intro}

In income tax systems, a ``pass-through'' entity (also known as a flow-through entity) refers to companies or corporations that are not subject to income taxes but whose income is ``passed through'' to their owners, who are taxed individually. Pass-through entities are very common in many countries. For example, in the USA, this type of firm (including sole proprietorships, general partnerships, limited partnerships, LLCs and S-corporations) increased from 83\% of all firms in 1980 to 94\% in 2007 \citep{CBOreport}. In particular, in Chile, a comprehensive tax reform was approved in 2014 that includes this type of firm and changes the tax basis to an accrual basis (``attributed income'') for both companies and individuals. More specifically, at the end of a calendar year, the taxable profits of a company would be attributed to the shareholders, proportionally to their participation. However, when a company or corporation incurs losses, they are not attributed to their shareholders but can be used as a credit for subsequent years. This is different from other countries, where losses are also passed through to the owners. 

A natural question for this taxation system is how to compute the final attributed income of each taxpayer. The difficulty of this comes from the fact that many companies and corporations are partially owned by other companies and corporations, successively constructing complex networks of companies, usually including cycles between them (i.e., a company can ``own'' a fraction of itself). A natural way to compute this is to iteratively distribute the positive income to shareholders and repeat this until all income has been assigned. However, many questions arise from this procedure: Is there relevance in the order in which companies are chosen to distribute at each step? Note that a company with negative income can receive sufficient attributed income to cover it losses and in a future iteration will start to distribute its received income to the shareholders. Does a unique final state exist for this system, independently of the order in which the incomes are attributed? Can we compute this final state efficiently? This article is motivated by a request of the Chilean taxation authority (Servicio de Impuestos Internos) to study all of these questions. 


In this paper, we mathematically formalize the problem and use the theory of Markov chains to prove that there exists a unique final state. We also prove that this state can be obtained by decomposing the network in a strongly connected component, in which the final attributed income can be computed efficiently. This leads to a fast algorithm to compute the final state, even for a large number of taxpayers. Finally, we show some results of the implementation of the algorithm on the real network of taxpayers in Chile.

To our knowledge, there is no such study in the literature. Nevertheless, similar questions are studied in the context of discrete games, particularly for chip firing games \citep{merino2005chip}. In the case of chip firing games over directed graphs \citep{bjorner1992chip}, each node contains a set of chips, and at each iteration, a node is selected and one chip is sent to each of its neighbors (if it has enough chips). The game stops if there exists no node with more chips than the number of its outgoing arcs. Note that multiple arcs are allowed between a given pair of nodes, so this game can be viewed as a discrete version of our problem when all starting incomes are positive. For this problem, the authors prove that the final state of the game (if it exists) is reached independently of the sequence of nodes chosen. Additionally, the authors remark that this problem can be seen as a tool for computing the absorption probabilities of certain Markov chains \citep{engel1975probabilistic}.

This paper is organized as follows. In Section \ref{concep}, we show the mathematical conceptualization of the problem, formulating the problem and defining the notation used in the subsequent sections. In Section \ref{calc}, we analyze how to calculate the attribution of income to shareholders, showing two particular cases first and then formulating the general case. In this section, we also show that the problem has only one finale state, which can be computed using the theory of absorbing Markov chains. In Section \ref{teoria}, we show the theoretical results needed to understand the validity of the algorithm that is explained in Section \ref{algoritmo} along with the two main proofs of this paper. In Section \ref{algoritmo}, we show the algorithm used to accurately and efficiently compute the attributed income. In addition, we prove the validity of this algorithm based on the theoretical results of the previous section. In Section \ref{resultados}, we analyze the network of Chilean taxpayers and the results of the algorithm, comparing its performance with that of alternative algorithms. Finally, we conclude and discuss the impact of this work on the Chilean taxation authority.


\section{Conceptualization}
\label{concep}

For simplicity, we refer to corporations as any pass-through entity and to individuals as any person or corporation that does not distribute its income.
Let $N$ be a set of taxpayers that consists of a subset of corporations $N_S$ and a subset of individuals $N_P$, such that $N_S \cup N_P=N$.
Each taxpayer $i\in N$ has an initial income $E^{(0)}_i$, which defines the vector $E^{(0)}$. Each corporation can be owned by corporations and individuals, represented by matrices $Q$ and $R$, respectively, where row $i$ represents the shares of corporations or individuals in corporation $i$. Therefore, $q_{ij}$ is the percentage of corporation $i$ owned by corporation $j$, and $r_{ij}$ is the percentage of corporation $i$ owned by individual $j$. Furthermore, we can assume that every individual is owned by himself. Thus, we define the matrix of shares $P$, where $p_{ij}$ is the percentage of taxpayer $i$ owned by taxpayer $j$. Matrix $P$ has the following form:

\begin{equation}\label{matriz_P}
P=\begin{bmatrix} Q&R \\ 0&I \end{bmatrix}.
\end{equation}

Moreover, $Q$ and $R$ have the following properties:
\begin{enumerate}
\item $0\leq Q \leq 1$ and $0\leq R \leq 1$,
\item $\lim_{n\to \infty} Q^n=0$,
\item $\sum_{j \in N} p_{ij}=1,\forall{i \in N}$.
\end{enumerate}
We denote by $\mathcal{P}$ the set of matrices $P$ with these properties.

\begin{remark}\label{obs1}
These properties ensure that every corporation is, directly or indirectly, owned by individuals.
\end{remark}

Given the vector of initial incomes $E^{(0)}$ and the matrix of shares $P$, we want to distribute the initial income to taxpayers and compute the attributed income of each taxpayer. A corporation with positive initial income must distribute its income proportionally to the share of each taxpayer. A corporation with negative initial income distributes its income only if the sum of its attributed income plus its initial income is greater than 0. Hence, for the distribution of income, a corporation with negative income would behave as an individual.

To consider corporations with negative income that do not distribute their income, we define the matrix of shares restricted in a subset of corporations as follows.
\begin{definition}
Let $S\subseteq N_S$ a subset of corporations; we define the matrix of shares restricted in $S$, named $P_S$, as:
\begin{equation}
\label{Def_1}
\left(P_{S}\right)_{i\bullet} =\begin{cases} P_{i\bullet} & \qquad \mbox{if } i \not\in S, \\
e_i & \qquad \mbox{if } i \in S, \end{cases}
\end{equation}
\end{definition}
where $e_i$ is the $i$-th canonical row vector.
Replacing row $i$ of $P$ by the $i$-th canonical row vector is equivalent to saying that corporation $i$ will not distribute its income (like individuals). In our case, we want to define a set $S$ that has all corporations with negative income. Thus, we use the following notation:
\begin{definition}\label{Def_2}
Given a vector of incomes $E$, we define a subset of corporations with negative income as
\[ S(E)= \{i\in N_s : E_i < 0 \}. \]
\end{definition}

Accordingly, we see that we can compute the attribution of income iteratively. In the first iteration, we distribute the initial income of each corporation proportionally to the share of each taxpayer, considering that corporations with negative initial income do not distribute. In the second iteration, we then compute the new income of each corporation, and these corporations distribute their new income to taxpayers again. In this second iteration, a corporation with negative initial income could now have positive income, so this corporation would distribute its income to its shareholders. The algorithm iterates until all income has been distributed to individuals or corporations with negative attributed income.

Using the notation that we introduced, this algorithm can be formalized as follows:

\begin{algo}\label{alg_1}
Let $E^{(n)}$ be the vector of attributed incomes in the $n$-th iteration of the algorithm, where $E_j^{(n)}$ is the income of taxpayer $j$. As in each iteration a corporation with negative attributed income does not distribute its income; the iteration can then be defined as
\begin{equation}
\label{ec_gen}
E^{(n+1)}=E^{(n)}P_{S^{(n)}},\ \text{ where } S^{(n)} = S(E^{(n)}).
\end{equation}
\end{algo}
We want to study whether $\lim_{n\to \infty} E^{(n)}$ exists and is unique, i.e.,, if the value of $E^{(n)}$ converges to a vector of final attributed income denoted by $E^{(\infty)}$.

Additionally, we can derive from equation \eqref{ec_gen} that if $E^{(\infty)}$ does exist, then
\begin{equation}\label{eq:extension}
E^{(\infty)}=E^{(0)}\cdot P_{S^{(0)}}\cdot P_{S^{(1)}}\cdot P_{S^{(2)}} \cdots
\end{equation}

It is important to note that $P_{S^{(n)}}$ is not necessarily different from $P_{S^{(n+1)}}$; indeed, $S^{(n)}$ probably does not change during several iterations. Moreover, there exists an iteration in which all corporations with negative income maintain negative income during all future iterations. Note also that a taxpayer with nonnegative income will remain nonnegative for all remaining iterations. These two observations imply that there is a final matrix $P_{S}$ that is multiplied infinite times to compute the vector of final attributed incomes. Therefore, there is a succession of integer numbers $n_1,\ldots,n_k$ such that the sets $S^{(n)}$ can be written as

\begin{equation*}
S^{(i)} = \begin{cases} S^{(n_1)} &i=0...n_1\\
S^{(n_j)} & i=n_{j-1}+1\ldots n_j, \ j=2\ldots k-1\\
S^{(n_k)} &i\ge n_{k-1}+1 \end{cases}.
\end{equation*}
Hence, equation \eqref{eq:extension} can be written as
\begin{equation}
\label{ec_gen_2}
E^{(\infty)}=E^{(0)}\cdot P_{S^{(n_1)}}^{n_1}\cdot P_{S^{(n_2)}}^{n_2-n_1}\cdot P_{S^{(n_3)}}^{n_3-n_2} \cdots P_{S^{(n_{k-1})}}^{n_{k-1}-n_{k-2}} \cdot P_{S^{(n_k)}}^\infty.
\end{equation}



The difficulty of the problem results from the fact that some corporations are owners of other corporations. If this does not occur, only one iteration would be necessary to compute the final attributed incomes of all taxpayers. 
In that case, the vector of final attributed income would be

\begin{equation*}
E^{(\infty)}=E^{(1)}=E^{(0)}P_{S^{(0)}}.
\end{equation*}

Because corporations are owners of other corporations, a corporation could be an owner of itself. This is why infinite iterations could be required, thus creating complex instances that could require a long time for computation.
However, as we show in Section \ref{calc}, we can use the theory of absorbing Markov chains to compute exactly the vector of final attributed income without iterating infinite times. This theory is the basis for designing the algorithm presented in Section \ref{algoritmo}.

\section{Computing the final attributed income}
\label{calc}

We start with two simple examples: 1) When all corporations have positive initial income and 2) When corporations with negative initial income have negative final attributed income.

\begin{example} \textit{Final attributed income when all initial incomes are positive}.

When all initial incomes are positive, all corporations distribute their income to their shareholders. Therefore, $S^{(n)}=\emptyset,\forall{n\ge0}$, which implies that $P_{S^{n}}=P,\forall{n\ge0}$; hence, we can see from equation \eqref{ec_gen_2} that
\begin{equation}
\label{all_+}
E^{(\infty)}=E^{(0)}P^{\infty}.
\end{equation}
\end{example}

\begin{example} \textit{Final attributed income when corporations with negative initial income have negative final attributed income}.

When corporations with negative initial income have negative final income, it is clear that $S^{(n)}=S,\forall{n\ge0}$, which implies that $P_{S^{(n)}}=P_{S^{(0)}},\forall{n\ge0}$, and we know from equation \eqref{Def_1} that the matrix $P_{S^{(0)}}$ is known and that $P_{S^{(0)}}\in \mathcal{P}$; therefore, we can see from equation \eqref{ec_gen_2} that

\begin{equation}
\label{some_very_-}
E^{(\infty)}=E^{(0)}P_{S^{(0)}}^{\infty}.
\end{equation}
\end{example}

In both cases, equations \eqref{all_+} and \eqref{some_very_-} require computation of $P^{\infty}$ and $P_{S^{(0)}}^{\infty}$, respectively, which has a closed formula, using a formula known from the absorbing Markov chains theory.

\subsection{Analogy with Absorbing Markov Chains}





An absorbing Markov chain has a set of transient states $N_S$ and a set of absorbing states $N_P$, defining a transition matrix $P$,
\begin{equation*}
P=\begin{bmatrix} Q&R \\ 0&I \end{bmatrix}
\end{equation*}
where $q_{ij}$ is the probability of moving from a transient state $i$ to a transient state $j$ in one time step and $r_{ij}$ is the probability of moving from a transient state $i$ to an absorbing state $j$ in one time step. At the same time, $0$ is a matrix of zeros, which means that the probability of moving from an absorbing state to a transient state is 0, and $I$ is the identity matrix, which implies that the probability of moving from an absorbing state $i$ to an absorbing state $j$ is 1 if $i=j$ and 0 otherwise.

The analogy between an absorbing Markov chain and our problem of attribution of incomes is evident. Taxpayers are defined as the states of the Markov chain, in which the percentages of each taxpayer owned by other taxpayers are represented as the transition probabilities of the Markov chain. Thus, corporations with negative initial incomes and individuals are defined as absorbing states, and the remaining corporations are defined as transient states. Moreover, it is known that a transition matrix of an absorbing Markov chain has the same properties as a matrix $P\in \mathcal{P}$. Thus, the matrices $Q$, $R$, $0$ and $I$ forming the matrix $P$ have the same properties and characteristics as the matrices of shares defined in equation \eqref{matriz_P}.

When all initial incomes are positive, the network of taxpayers is modelled as an absorbing Markov chain, in which each individual is an absorbing state and each corporation is a transition state. Conversely, when all corporations with negative initial income have negative final attributed income, corporations with positive initial income remain as transition states, and individuals remain as absorbing states, but corporations with negative initial income are absorbing states.

The Chapman–Kolmogorov equation of Markov chains defines the probability of moving from a state $i$ to a state $j$ in $k$ time steps as the component $(i,j)$ of the matrix $P^k$. It is known that in an absorbing Markov chain, when $k$ tends to infinity, the process will be absorbed by one state and remain there forever. Therefore, when there are multiple absorbing states, the question is to compute the probability of being absorbed by a certain state. These are called absorption probabilities, which are precisely $P^\infty=\lim_{k\to \infty} P^k$, where $(P^\infty)_{ij}$ is the probability of being absorbed by an absorbing state $j$ given that the state of the Markov chain is $i$. Indeed, $P^\infty$ can be computed as

\begin{equation} \label{P_infinity}
P^\infty= \begin{bmatrix} Q^\infty &(\sum_{i=0}^\infty Q^i) \cdot R \\ 0&I \end{bmatrix} = \begin{bmatrix} 0&(I-Q)^{-1}R \\ 0&I \end{bmatrix}
\end{equation}

Extending the analogy of absorbing Markov chains with the problem shown in this work, component $(i,j)$ of matrix $P^k$ is the percentage of income of taxpayer $i$ distributed to taxpayer $j$ after iteration $k$ with Algorithm \ref{alg_1}. Similarly, the percentage of income of taxpayer $i$ attributed to taxpayer $j$ is component $(i,j)$ of matrix $P^\infty$. Therefore, using formula \eqref{P_infinity}, we can compute the final attributed income of each taxpayer in both particular cases shown in Examples 1 and 2.

\subsection{General case}
In general, some taxpayers with negative initial income have positive final attributed income. Equation \eqref{ec_gen_2} defined that
\begin{equation*} 
E^{(\infty)}=E^{(0)}\cdot P_{S^{(n_1)}}^{n_1}\cdot P_{S^{(n_2)}}^{n_2-n_1}\cdot P_{S^{(n_3)}}^{n_3-n_2} \cdots P_{S^{(n_{k-1})}}^{n_{k-1}-n_{k-2}} \cdot P_{S^{(n_k)}}^\infty,
\end{equation*}
where $E^{(0)}$ is the vector of initial incomes, $P_{S^{(n_i)}}$ are existent and unique matrices $\forall i=1,\ldots,k$ and, as we can see in equation \eqref{P_infinity}, $P_{S^{(n_k)}}^{\infty}$ exists and is unique. Therefore, the vector of final attributed incomes $E^{(\infty)}$ exists and is unique. This ensures that given a vector $E^{(0)}$ and a matrix of shares $P\in \mathcal{P}$, if we iterate with Algorithm \ref{alg_1}, we have only one final distribution of the initial income to individuals.

However, as we show in the next section, it is unnecessary to distribute the income of all corporations simultaneously. Indeed, if we distribute only the income of arbitrary subsets of corporations (keeping in mind that a corporation with negative income does not distribute), the vector of final attributed incomes is the same as that computed using Algorithm \ref{alg_1}.

\section{Mathematical properties of the matrix of shares}
\label{teoria}

The next lemma shows that it is unnecessary to iterate infinitely, and the only reason to compute the attributed income is to know which corporations will have negative final attributed income $S^{(\infty)}=\{i\in N_S:E_{i}^{(\infty)}<0\}$.

\begin{lemma} \label{lem:base}
Let $P\in \mathcal{P}$ be a matrix of shares with a set of corporations $N_S$. Then, for any subset of corporations $S\subseteq N_S$
\begin{equation*}
P_S \cdot P^{\infty}=P^{\infty}. 
\end{equation*}
\end{lemma}
Note that this lemma implies that $P_S^k P^\infty = P^\infty$ for any $k=1\ldots \infty$. This will be key for the algorithm proposed in the next section.


\ifopre
\proof{Proof.}
\else
\begin{proof}
\fi
By equation \eqref{Def_1}, the $i$-th row of $P_S \cdot P^\infty$ in the case in which $i\notin S$ is given by
\[ \left(P_S\cdot P^{\infty}\right)_{i \bullet} = (P_S)_{i \bullet} \cdot P^{\infty} = P_{i \bullet} \cdot P^{\infty} = P^{\infty}_{i \bullet} \]
where the last equality is given because $P\cdot P^{\infty}=P^{\infty}$. In contrast, if $i\in S$, then
\[ \left(P_S\cdot P^{\infty}\right)_{i \bullet} = (P_S)_{i \bullet} \cdot P^{\infty} = e_i \cdot P^{\infty} = P^{\infty}_{i \bullet} \]
proving the result.
\ifopre
\Halmos\endproof
\else
\end{proof}
\fi

The foregoing lemma says that if in one iteration we do not attribute income for a subset of corporations $S$, and in the following iterations we attribute income as usual with matrix $P$, then the final attributed income will be the same as if we do not skip any iterations.
Moreover, because for all steps $i$, the subset $S^{(n_i)}$ contains the last subset $S^{(n_k)}$, Lemma \ref{lem:base} says that 
\begin{equation*} 
P_{S^{(n_1)}}^{n_1}\cdot P_{S^{(n_2)}}^{n_2-n_1}\cdot P_{S^{(n_3)}}^{n_3-n_2} \cdots P_{S^{(n_{k-1})}}^{n_{k-1}-n_{k-2}} \cdot P_{S^{(n_k)}}^\infty=P_{S^{(n_k)}}^\infty \label{eq:proc}
\end{equation*}
Moreover, it is also true that
\begin{equation*} 
P_{S^{(n_1)}}^{\infty}\cdot P_{S^{(n_2)}}^{\infty}\cdot P_{S^{(n_3)}}^{\infty} \cdots P_{S^{(n_{k-1})}}^{\infty} \cdot P_{S^{(n_k)}}^\infty=P_{S^{(n_k)}}^\infty. \label{eq:procinfty}
\end{equation*}
Therefore, from equation \eqref{ec_gen_2}, the vector of final attributed incomes can be computed as
\[ E^{(\infty)} = E^{(0)} P_{S^{(n_k)}}^\infty \]
where $S^{(n_k)}=S(E^{(\infty)})$.

This property says that if we are able to guess the subset of corporations that will finish with negative attributed income, we require only one step to find this final state. Unfortunately, it is impossible to know a priori the set $S(E^{(\infty)})$ of corporations that finish with negative attributed income from the initial information $P$ and $E^{(0)}$.
However, the next theorem shows that when starting to distribute an arbitrary subset of corporations in each iteration (keeping in mind that a corporation with negative attributed incomes does not distribute) and then returning to the usual iterations, the algorithm still converges to the vector of final attributed incomes $E^{(\infty)}$.

\begin{theorem}\label{teo:principal}
Let $\{\hat{E}^{(j)}\}_{j\geq 0}$ be a sequence of vectors of incomes such that
\[ \hat{E}^{(0)}=E^{(0)} \quad \text{ and } \quad \hat{E}^{(j)} = \hat{E}^{(j-1)} P_{T^{(j-1)}} \quad \text{for } j=1\ldots k\] 
where $T^{(0)},\ldots,T^{(k-1)}$ are subsets of corporations such that $T^{(j)} \supseteq S(\hat{E}^{(j)})$, and let
\[ \hat{E}^{(j)} = \hat{E}^{(j-1)} P_{S(\hat{E}^{(j-1)})} \quad \text{for }j>k\] 
then,
\[ \lim_{j\to\infty} \hat{E}^{(j)} = E^{(\infty)}. \]
\end{theorem}

\ifopre
\proof{Proof.}
\else
\begin{proof}
\fi
Note that if $\hat{E}^{(k)}$ satisfies $S(\hat{E}^{(k)})\supseteq S(E^{(\infty)})$, then from Lemma \ref{lem:base}, we know that
\begin{equation} \label{eq:arg_teo}
P_{T^{(0)}} \cdots P_{T^{(k-1)}}\cdot P^\infty_{S(E^{(\infty)})} = P^\infty_{S(E^{(\infty)})} 
\end{equation}
Therefore, $\hat{E}^{(\infty)}=E^{(0)}\cdot P^\infty_{S(E^{(\infty)})} = E^{(\infty)}$. 

Let us assume that there is an $i^*\in S(E^{(\infty)})$ such that $\hat{E}^{(k)}_{i^*} > 0$. Without loss of generality, we can assume that this is the first iteration in which $i^*$ exists. Therefore, $S(E^{(\infty)}) \nsubseteq S(\hat{E}^{(k)})$ but $S(E^{(\infty)}) \subseteq S(\hat{E}^{(k-1)})$. In this case, equation \eqref{eq:arg_teo} is still true, and
\[ \underbrace{E^{(0)}\cdot P_{T^{(0)}} \cdots P_{T^{(k-1)}}}_{\hat{E}^{(k)}}\cdot P^\infty_{S(E^{(\infty)})} = E^{(0)}\cdot P^\infty_{S(E^{(\infty)})} = E^{(\infty)} \]

However, this is impossible because $\hat{E}^{(k)}_{i^*}>0$ and $i^*\in S(E^{(\infty)})$, so $i^*$ does not distribute its income, and component $i^*$ of $\hat{E}^{(k)}\cdot P^\infty_{S(E^{(\infty)})}$ must be greater than 0, resulting in a contradiction.
\ifopre
\Halmos\endproof
\else
\end{proof}
\fi

This theorem implies that if we start from $E^{(0)}$, we are able to find an income vector $\bar{E}$ such that $\bar{E}_i\leq 0$ $\forall i\in N_S$, using subsets $T^{(0)},\ldots,T^{(k-1)}$ with the properties noted in the theorem (even if we iterate infinite times with a matrix $P_{T^{i}}$), $\bar{E}$ is the vector $E^{(\infty)}$, because $\bar{E} P_{S(\bar{E})} = \bar{E}$.
This assertion is the key to understanding the validity of the algorithm proposed in the next section.

\section{Algorithms for computing the final attributed income}
\label{algoritmo}

Recall that Algorithm \ref{alg_1} could require an infinite number of iterations to converge to the final attributed income, which is not allowable. This could be limited by iterating until a small amount has not been attributed but still could lead to many iterations.
A different way to compute the vector of final attributed incomes results from Lemma \ref{lem:base} and equation \eqref{ec_gen}, by iteratively computing
\[ E^{(n+1)}=E^{(n)}P^\infty _{S^{(n)}},\ \text{ where } S^{(n)} = S(E^{(n)}). \]
This procedure will be completed in no more than $|N_S|$ iterations. 
However, from the computational point of view, it is costly to invert a large matrix. Specifically, inversion of a dense matrix of size $k\times k$ has a computational complexity of $\mathcal{O}(k^{2.3728})$ \citep[see][]{le2014powers,coppersmith1987matrix}. Nevertheless, most software, including the de facto standard library LAPACK~\citep{laug}, implements a classic $\mathcal{O}(k^3)$ algorithm. Hence, the proposed algorithm to solve the problem can be computationally intractable for a large number of corporations $N_S$.
Theorem \ref{teo:principal} proves that we can decompose the problem into smaller subproblems to obtain the final attributed income. A natural way to decompose the problem is by using strongly connected components. Given the participation matrix $P$, we can define a directed graph $G=(V,A)$ in which each vertex $v \in V$ is a taxpayer, and we add an arc $(u,v)\in A$ if and only if $p_{uv}>0$. A strongly connected component on this graph represents a subset of corporations in which all are indirectly owned by themselves. It is a well-known property that any directed graph can be decomposed in a set of strongly connected components such that if we contract each strongly connected component into a vertex, we obtain a directed acyclic graph \citep[see][]{bang2008digraphs}. Moreover, Tarjan's algorithm \citep{tarjan1972depth} allows the decomposition of the graph into strongly connected components and returns an acyclic ordering of them, in time $\mathcal{O}(|V|+|E|)$.

Hence, we can apply this decomposition to solve our problem more efficiently. Given an acyclic ordering of the strongly connected components, we can compute the attributed income of each taxpayer in a given component and distribute it to the corresponding shareholders in other strongly connected components. Because this is an acyclic ordering, taxpayers in the initial component will not receive further incomes, so the obtained attributed income will be definitive.

\algrenewcommand\algorithmicrequire{\textbf{Input:}}
\algrenewcommand\algorithmicensure{\textbf{Output:}}
\begin{algorithm}[hbtp]
\caption{Compute the attributed income of each taxpayer} \label{alg:main}
\begin{algorithmic}[1]
\Require Taxpayers $N_S,N_P$, participation matrix $P$, initial income $E^{(0)}$
\Ensure Attributed income $E$
\State $E \gets E^{(0)}$
\State Execute Tarjan's algorithm to compute an acyclic ordering of the strongly connected components.
\For{each strongly connected component $C$} \Comment{in the acyclic order previously computed}
\If{$C=\{v\}$} \Comment{Component has only one vertex}
\If{$E_v > 0$} \Comment{The taxpayer has positive income}
\For{ each shareholder $u$} \Comment{Distribute its income among the shareholders}
\State $E_u \gets E_u + E_v \cdot p_{uv}$ 
\EndFor
\State $E_u \gets 0$
\EndIf
\Else \Comment{Component has at least two corporations}
\Repeat
\State $Redo \gets 0$
\State $S \gets N_S \setminus \{v\in C : E_v > 0\}$ \Comment{Consider only corporations in $C$ with positive income}
\State $E \gets E \cdot P^{\infty}_S$ \Comment{Distribute its incomes among shareholders}
\If{exist $v\in C\bigcap S$ such that $E_v>0$} 
\State $Redo \gets 1$ \Comment{A corporation with negative income now has positive income}
\EndIf
\Until{$Redo = 0$}
\EndIf
\EndFor
\end{algorithmic}
\end{algorithm}

Pseudocode of the proposed algorithm is presented in Algorithm \ref{alg:main}. Note that for each strongly connected component $C$, to obtain the final attributed income of its taxpayers, it requires inversion of a matrix of size no more than $|C|$, which is repeated each time a taxpayer in $C$ with negative income finishes with positive income. Hence, the computational complexity to obtain the attributed income of a component requires no more than $|C|$ inversions of a matrix of size $|C|$. Hence, it will be appropriate to use this algorithm only if the size of the strongly connected components is small, which is expected in a network of corporations, as we will see in the following section.

At the end of the algorithm, the attributed income $E$ will be nonpositive for the corporations, and nonnegative for the individuals. Hence, our final attributed income $E$ satisfies that $E=E\cdot P_{S(E)}$, so by Theorem \ref{teo:principal} and its following discussion, the obtained income is effectively the vector of final attributed incomes $E^{(\infty)}$.

\section{Computational Experiments and Data Analysis}
\label{resultados}

We have presented an algorithm to obtain the final attributed incomes and we showed that its performance is given by the size of its connected components, so strongly depending on the topological characteristic of the network. Therefore, in this section we provide an analysis of the Chilean taxpayer network to understand its characteristics, and we benchmark our algorithm on this network.

The data described in this section are provided by the Chilean taxation authority and correspond to the $2015$ fiscal year.
The original taxpayers' network consists of $1\,240\,809$ individuals and $786\,293$ corporations, which are connected by $2\,568\,182$ links of which $\approx 90\%$ connect corporations with individuals. However, we can simplify this network by removing trivial components, which consist of one corporation that owns nothing and is owned only by individuals. These components are trivial because the distribution of income can be solved in one iteration of Algorithm \ref{alg_1}. After this simplification, the new taxpayer network consists of $356\,372$ individuals and $152\,914$ corporations connected by $1\,122\,875$ links of which $\approx 75\%$ connect corporations with individuals.

In the simplified network, corporations own an average of $1.78$ corporations, and the average quantity of owners per corporation is $7.23$, of which $5.48$ are individuals and $1.75$ are corporations. Moreover, $40\%$ and $34\%$ of corporations have in-degrees of $0$ and $1$, respectively, and $45\%$ of corporations have an out-degree of $2$. Moreover, the average number of corporations owned per individual is $2.39$, among which $56\%$ and $20\%$ of individuals have in-degrees of $1$ and $2$, respectively.

\begin{figure}
\ifopre
\else
\centering
\includegraphics[width=.7\linewidth] {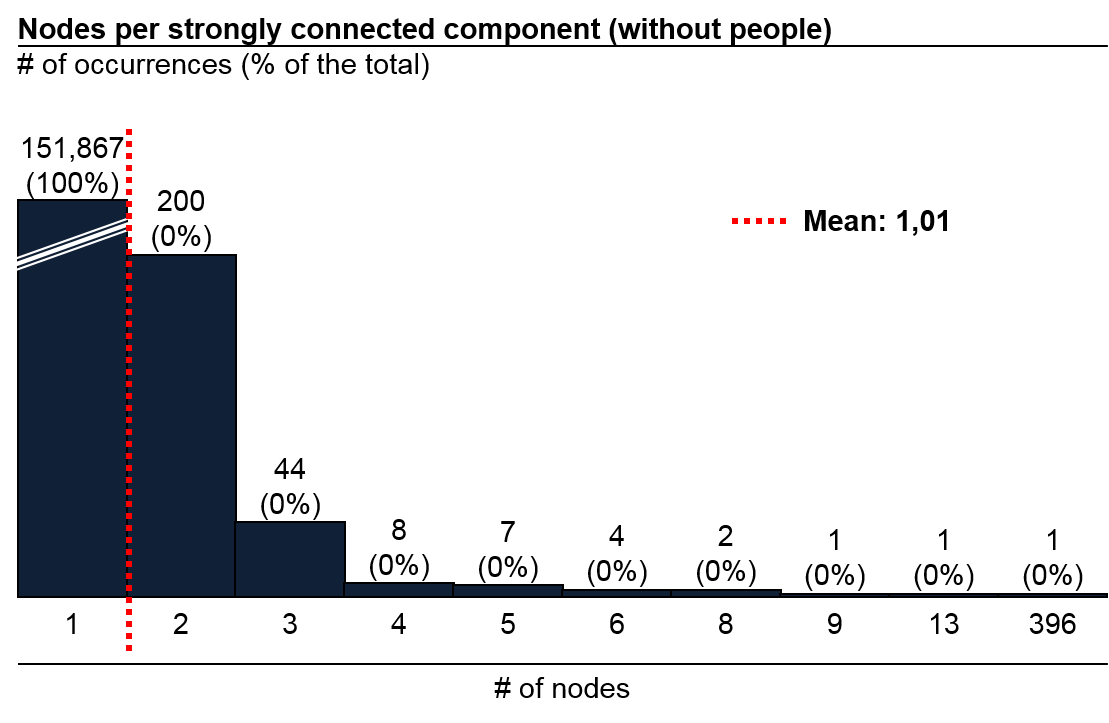}
\caption{Size of strongly connected components} \label{fig_hist_str_com}
\fi
\end{figure}

The complexity of the problem is given by corporations, not individuals. Therefore, we analyze the corporation network by removing individuals. This network has $152\,135$ strongly connected components, of which only $268$ contain more than one node. As we can see in Figure \ref{fig_hist_str_com}, $200$ of these $268$ components consist of $2$ nodes, and the largest component consists of $396$ nodes. Therefore, using the algorithm described in this paper, the largest matrix that we must invert is a $396 \times{} 396$ matrix.

\begin{figure}[tbp]
\ifopre
\else
\centering
\includegraphics[width=.48\linewidth] {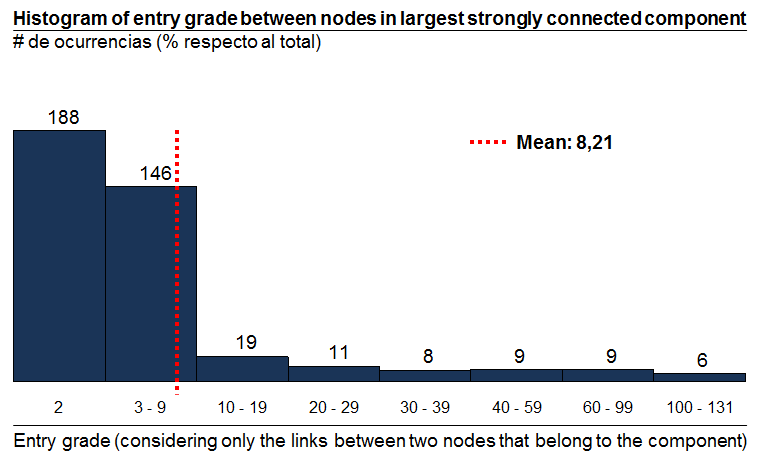}
\includegraphics[width=.48\linewidth] {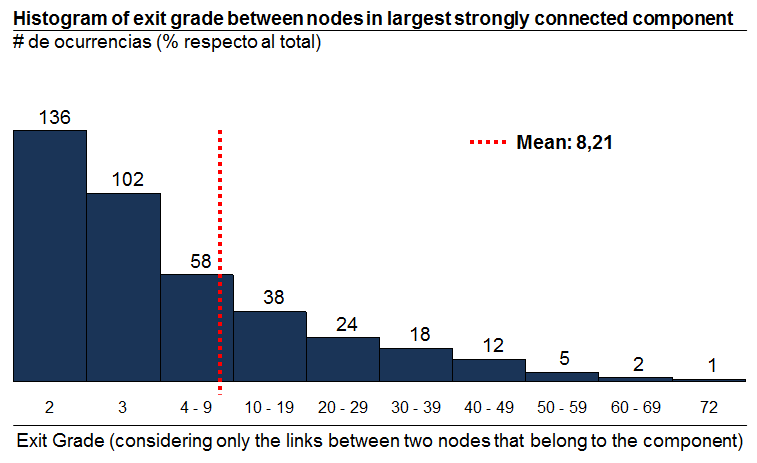}
\caption{In-degree and out-degree of nodes in largest strongly connected component} \label{fig_hist_str_com_large_entry}
\fi 
\end{figure}

\begin{figure}[tbp]
\ifopre
\else
\centering
\includegraphics[width=.5\linewidth] {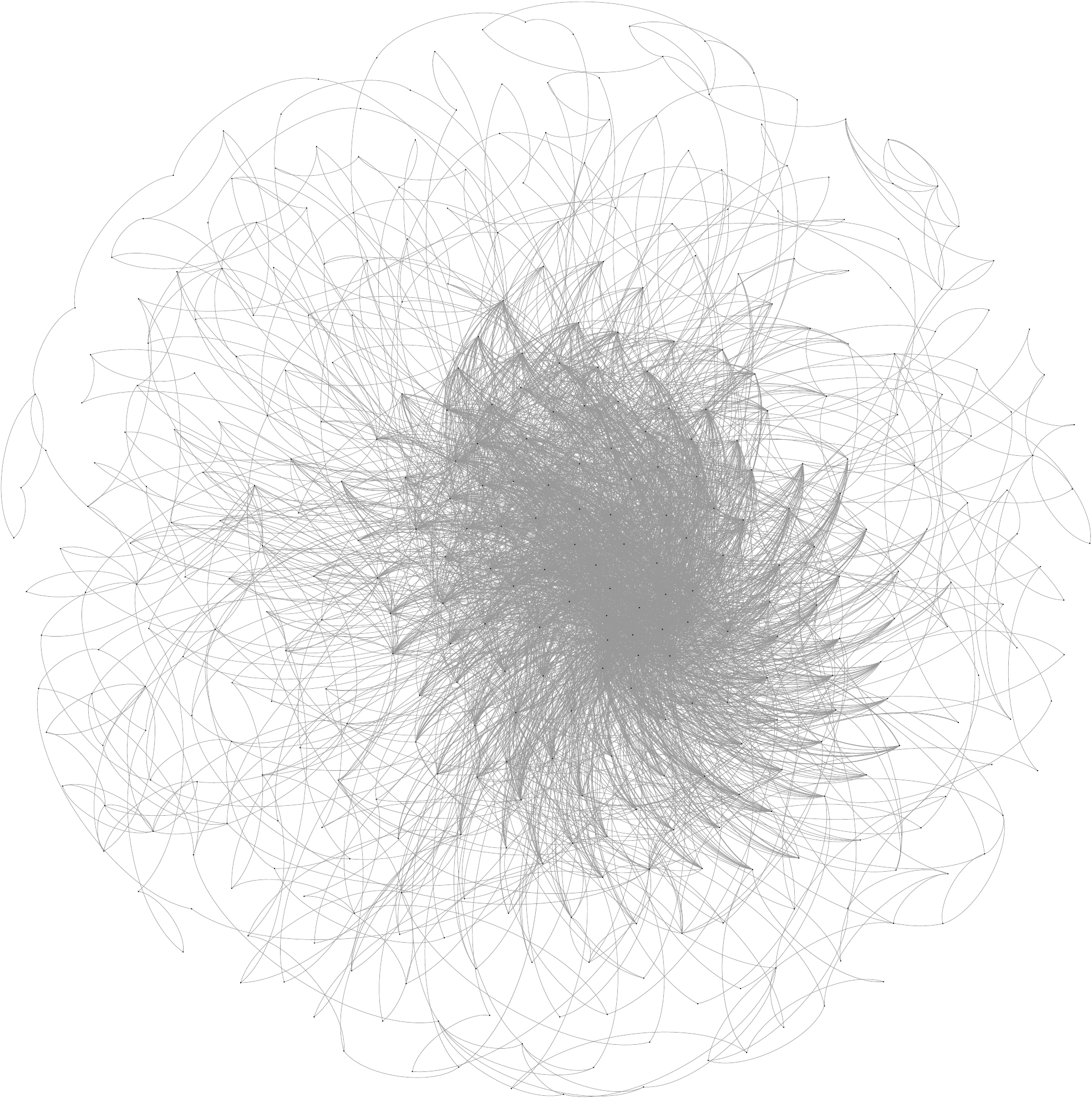}
\caption{Diagram of the largest strongly connected component} \label{fig_red_str_com_large}
\fi
\end{figure}

As we can see in Figure \ref{fig_hist_str_com_large_entry} and Figure \ref{fig_red_str_com_large}, the corporations inside the largest strongly connected component are highly connected. Indeed, the average out-degree and in-degree of the nodes of this component are both $8.21$ (considering only the links between two nodes that belong to the component); there are $10$ corporations directly connected to more than $100$ other nodes of the component, and more than $10\%$ of nodes are directly connected to at least $40$ nodes.

Although there is only one complex strongly connected component, many strongly connected components are weakly connected, creating a large weakly connected component of $90\,322$ strongly connected components for a total of $91\,011$ corporations (if we do not remove individuals for the analysis of weakly connected components, the largest weakly connected component would connect $462\,649$ taxpayers, more than $90\%$ of the nodes of the simplified network). This adds complexity to the problem because the algorithm must respect the precedence of the strongly connected components. However, $81\%$ of the weakly connected components consist of no more than $3$ corporations.

The algorithm described in this paper was implemented using the C programming language, using the libraries BLAS/LAPACK to invert the matrices. Note that even if matrices $Q$ are sparse, the inverse of $(I-Q)$ is not necessarily sparse, so memory must be allocated for the whole matrix $Q$. This makes it impossible to solve the problem using equation \eqref{ec_gen_2} because $Q$ has size $152\,914 \times 152\,914$, requiring more than 150 GB of RAM to invert just this matrix. 

We implemented Algorithm \ref{alg:main} for this data instance. The algorithm required 344 matrix inversions to compute the final attributed incomes. The complete algorithm runs in less than 10 s. As a comparison, an implementation of Algorithm \ref{alg_1} iterated until the maximum income of a corporation is less than \$1 requires several hours to finish.

\section{Conclusions}

Using an analogy to Markov chains, we construct an efficient algorithm to compute the final attributed income of taxpayers in a pass-through tax system. The complexity of the algorithm is based on the size of the largest strongly connected component of the taxpayer network. We also prove that this final income is unique and robust to any order in which corporations attribute their incomes. This fact allow us to decompose the problem into strongly connected components, which can be obtained using Tarjan's algorithm. The decomposition is the key property of the proposed algorithm, allowing us to solve large-scale networks in a few seconds.

An algorithm that compute income taxes sufficiently fast could lead to the ability to perform further analysis for a taxation authority, such as evaluating the impact of having a mixed system with entities allowed to choose whether to attribute their incomes, forecasting tax collection for future years, to evaluate the impact of tax exemptions, or simply computing the income obtained from different types of entities (e.g., foreign companies). In a more general setting, for any country (with or without pass-through entities), this methodology allows a better estimation of the distribution of wealth to be obtained for the richest deciles, which are usually underestimated in inquiry-based studies~\citep{EstudioBancoMundial}. 

\ifopre
\else
\subsubsection*{Acknowledgments} The authors gratefully acknowledge the Department of Studies, Servicios Impuestos Internos, particularly to Carlos Recabarren, for introducing us the problem and its relevance, and their valuable collaboration that lead us to obtain these results.
\fi

\bibliographystyle{ormsv080}
\bibliography{sii}






\end{document}